\documentclass[12pt, draftcls,onecolumn]{article}

\usepackage{acronym}
\usepackage[utf8]{inputenc}
\usepackage{graphicx} % Required for inserting images
\usepackage{subcaption}
\usepackage[colorlinks=true, linkcolor=black, citecolor=red]{hyperref}
\usepackage{amsmath} % For advanced math environments like 'aligned'
\usepackage{amssymb} % For math symbols like \mathbb
\usepackage{algorithm} % For the algorithm float environment
\usepackage{algpseudocode} % For basic pseudocode structures like \State and \For
\usepackage{hyperref}
\usepackage{authblk}
\usepackage{float} % For forcing figures positions
\usepackage{mathtools}
\usepackage{amsthm} % Provides theorem-like environments, including the proof environment
\usepackage{enumitem}
\usepackage[svgnames]{xcolor} % For coloring text, including named colors like ForestGreen

\theoremstyle{remark}
\newtheorem*{remark}{Remark}

\graphicspath {{./Network/} {./FOLDER2/}} 

\newcommand{\keywords}[1]{\textbf{Keywords:} #1} % For Keywords

\begin{document}
\title{Smart nudging for efficient routing through networks}

% \author{Pouria M. Oqaz, Emanuele Crisostomi, Elena Dieckmann,\\
% Robert Shorten}

\author[1]{Pouria M. Oqaz}
\author[2]{Emanuele Crisostomi}
\author[1]{Elena Dieckmann}
\author[1]{Robert Shorten}

\affil[1]{\small{Dyson School of Design Engineering, Imperial College London, UK}}

\affil[2]{\small{Department of Energy, Systems, Territory and Constructions Engineering, University of Pisa, Italy}}

\date{}
\maketitle

\begin{abstract}

In this paper, we formulate the design of efficient digitalised deposit return schemes as a control problem. We focus on the recycling of paper cups, though the proposed methodology applies more broadly to reverse logistics systems arising in circular economy R-strategies. Each item is assumed to carry a digital wallet through which monetary rewards are allocated to actors transferring the item across successive stages, incentivising completion of the recycling process. System efficiency is ensured by: (i) decentralised algorithms that avoid congestion at individual nodes; (ii) a decentralised AIMD-based algorithm that optimally splits the deposit across layers; and (iii) a feedback control loop that dynamically adjusts the deposit to achieve a desired throughput. The effectiveness of the framework is demonstrated through extensive simulations using realistic paper cup recycling data.\\

\end{abstract}

\section{Introduction}
\label{sec:introduction}

\subsection{Motivation and State of the Art}
\label{sec:intro:motivation}

The circular economy represents a fundamental shift from the traditional linear model of “take–make–waste” toward a regenerative system designed to minimise waste and optimise resource efficiency~\cite{kirchherr2023}. By prioritising R-strategies such as \textit{reuse, repair, remanufacturing, and recycling,} the circular economy seeks to maintain the value of materials and products for as long as possible, thereby reducing the environmental pressures associated with resource extraction and waste generation~\cite{R_strategies_2006}. This model not only supports ecological sustainability but also offers substantial economic opportunities and stimulates innovation in product design and supply chain management~\cite{CE_PD_2023}. To fully realise its potential, however, coordinated policy actions, cross-sector collaboration, and significant investments in infrastructure and research are required~\cite{cE_CrossSectoral}.\\

Deposit Return Schemes (DRS), also known as Deposit Refund Systems, play a critical role in advancing the circular economy by creating effective incentives for the collection and high-quality recycling of beverage containers~\cite{Review_DRS_Potential_2025}. By refunding a small deposit when consumers return used packaging, these systems significantly increase recovery rates, reduce litter, and ensure a consistent supply of clean, recyclable materials for closed-loop production~\cite{Calabrese_2025}. DRS not only support resource efficiency and waste reduction but also encourages responsible consumer behaviour and fosters collaboration among producers, retailers, and recycling operators~\cite{Kukenthal2023}.\\

While in traditional DRS implementations, a single refund is provided only to the consumer once the product or its packaging is returned, more recent developments extend these single-refund schemes toward digitalised and multi-actor variants in which rewards may be distributed to all participants contributing to the reverse-logistics chain~\cite{gong2022blockchain, LU2022100048, Tang18062021}. References~[8–17], for instance, explore Internet-of-Things (IoT) enabled, blockchain-supported, and gamified architectures for incentivised recycling.\\

In particular, in 2020, \cite{10.1007/978-3-030-32523-7_72} introduced a distributed-ledger cyber-physical architecture in which digital tokens are bound to individual paper cups to enable a refundable deposit mechanism that enforces responsible disposal. In 2021, \cite{ZHOU2021142} designed a smart, data-driven incentive-based recycling system integrating IoT monitoring, pattern discovery, dynamic price adjustment, short-term forecasting, and multi-stakeholder information-sharing to improve household collection. In 2022, \cite{gong2022blockchain} developed a blockchain-enabled recycling framework that uses distributed-ledger traceability, smart contracts, and token incentives to coordinate recycling activities and enhance system performance. In 2024, \cite{doi:10.1177/0734242X241296617} analysed how modern DRS implementations strengthen circular economy outcomes by increasing return rates and enabling high-quality closed-loop recycling, while \cite{lakhan2024evaluating} evaluated global DRS effectiveness and showed that well-designed schemes routinely exceed 90 per cent return rates but face economic and operational constraints. More recently, in 2025, \cite{Mills20012025} examined how gamified digital platforms use points, badges, and lottery-style rewards to incentivise pro-environmental behaviour, \cite{ibrahim_smart_2025} proposed an IoT-enabled and behaviour-informed plastic-recycling system using customised incentives to increase urban participation, and \cite{doi:10.4491/eer.2024.241} reviewed global glass-recycling systems, highlighting how DRS, Extended Producer Responsibility (EPR), and digitalisation can enhance collection rates and support circular economy goals.\\

A representative example of modern digitalised {DRS} (i.e., {DDRS}) is the IOTA “KupKrush” prototype,\footnote{\href{https://biotasphere.com/}{https://biotasphere.com/}} which demonstrates how digital identities and automated micropayments may support material traceability and multi-layer rewards along the recycling process.\\

However, such schemes remain largely conceptual and do not address several fundamental {challenges spanning operational, architectural, and incentive-design aspects}: (i) the absence of a mechanism to prevent congestion at smart recycling bins {, because in contrast to household-based DRS models \cite{Eunomia2023SerialisedDDRS, PolytagOcado2023}, IOTA KupKrush relies on physical smart bin infrastructure, where limited storage capacity can lead to congestion and load imbalance \cite{Walk2020HalfEmpty, Bano2020AIoTSmartBin, Abuga2021SmartGarbageBin} ;} (ii) the lack of a decentralised method to split a fixed deposit among heterogeneous participants; and (iii) the absence of a dynamic strategy for adjusting the deposit level in response to real-time recycling performance. {The last aspect is particularly critical: indeed, the effectiveness of DRS heavily relies on the calibration of the deposit value. Insufficient deposit levels may result in weak consumer incentives for recycling, while excessively high deposits may cause disproportionate price increases, negatively affecting product demand.}\\

These limitations highlight the need for a control- and optimisation-based framework that supports decentralised decision-making, real-time feedback, and scalable coordination across digitally incentivised reverse-logistics networks.\\

\subsection{Contributions and Organisation}
\label{sec:intro:contributions}

This paper complements and completes a {DDRS} framework by specifically addressing the three major gaps identified above in the KupKrush protocol. Our main contributions are:
\begin{enumerate}
    \item \textbf{Decentralised load balancing:}  
    We design a probabilistic, decentralised assignment algorithm in which smart recycling bins broadcast Poisson-distributed signals whose frequency depends on their occupancy. This induces users to select less congested bins, achieving balanced utilisation similar performance to a centralised solution.

    \item \textbf{Decentralised reward allocation via extended AIMD:}  
    We formulate reward-splitting as a constrained optimisation problem with a multiplicative utility structure and develop an extension of the classical {AIMD} algorithm that converges to the optimal reward allocation in a decentralised and privacy-preserving manner.

    \item \textbf{Dynamic deposit pricing:}  
    We introduce a {PID}-based feedback loop that adjusts the deposit in real time so that the observed recycling rate tracks a desired target. {In doing so, we avoid recommending sudden changes in the value of the deposits, to allow users to slowly adjust their recycling habits, through small, gradual dynamic nudges}.
\end{enumerate}

Although paper cup recycling serves as our running example, the proposed framework applies broadly to other reverse logistics and multi-layer recovery processes.\\

\noindent
\textbf{Organisation:}  
Section~\ref{sec:problem} formalises the system model and problem definitions.  
Section~\ref{sec:load_balancing} presents the decentralised load balancing algorithm.  
Section~\ref{sec:AIMD} develops the extended {AIMD} algorithm for reward allocation.  
Section~\ref{sec:PID} introduces the PID-based deposit controller.  
Section~\ref{sec:con} concludes the paper.

\section{Problem Statement}
\label{sec:problem}

\begin{figure}[H]
    \centering
    \includegraphics[width=1\linewidth]{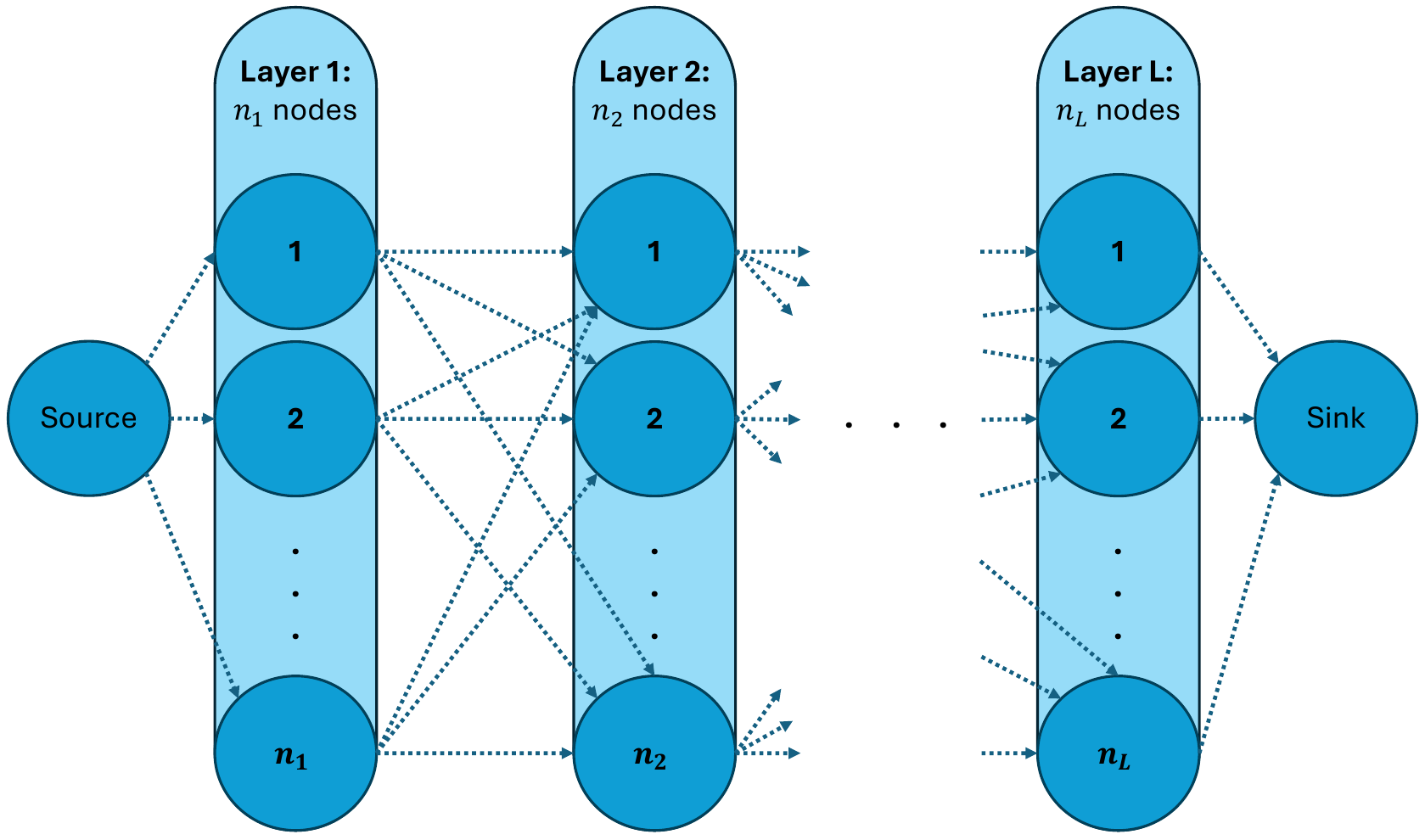}
    \caption{\small A network model of a reverse logistics supply chain represented as a directed acyclic graph. The graph starts at a source node, ends at a sink node, and includes $L$ intermediate layers. Each layer $i$ contains $n_i$ nodes. For instance, in the context of a recycling network, nodes in the first layer represent recycling bins, nodes in the second layer represent collection centres, and so on until nodes in the last layer represent paper mill factories.}
    \label{fig:Network}
\end{figure}

We consider a specific form of reverse logistics network that is monetarily incentivised, aiming at correctly moving the item from source to its final destination sink (e.g., recycling a used paper cup), as depicted in Fig.~\ref{fig:Network}.

For instance, a simplified reverse logistics chain for the recycling of paper cups may consist of three layers: \textit{(i)} the first layer consists of smart recycling bins, and people are responsible for duly throwing away their used paper cups into such devices; \textit{(ii)} the second layer consists of recycling collection centres. They are the participants that move paper cups from the first to the second layer; \textit{(iii)} finally, paper mills may use recycled paper cups from the second layer to produce new paper sheets.\\

In this work, following the KupKrush model, we assume that a paper cup carries a digital monetary deposit, and a portion of this deposit is transferred to participants each time the cup progresses from one layer to the next. Accordingly, participants are incentivised to actually support such transitions and thus to actively participate in the recycling process, as in doing so, they can be monetarily rewarded.\\

In particular, in this supply chain, whoever transfers the object from layer $i$ to the next layer $i+1$ receives a portion of the deposit $D$ as their reward $R_{i+1}$. For example, when coffee drinkers throw the used paper cup in a smart recycling bin, they receive £ $R_1$; or when waste collectors collect paper cups from smart recycling bins, they receive £ $R_2$ as their monetary reward. Since all of the rewards are going to be paid from the deposit, we have that

\begin{equation}
    \sum_{i=1}^L R_i \leq D,
    \label{eq:R_i<D}
\end{equation}
where $i$ is the index denoting the layer; $L$ is the total number of layers in the network; $R_i$ is the reward associated with layer $i$; $D$ is the initial deposit in the digital wallet of the object, usually paid by the consumer (e.g., a coffee drinker in our application). {The inequality \ref{eq:R_i<D} is always valid throughout this manuscript.}

In this manuscript, we address three open problems related to the KupKrush model that we have just presented.

\subsection{Congestion and Load Balancing}
\label{Problem_1}

In the KupKrush model, the problem of congestion arises when one of the nodes in a layer is full, and can not accommodate further material from the previous layer, until it is appropriately emptied~\cite{9451652}. For instance, a person may go to a smart recycling bin to throw away a paper cup, only to find out that the bin is already full. Clearly, this may discourage the customer, who may eventually throw away the paper cup in a conventional bin and decline the reward. This problem is particularly relevant since the number of smart recycling bins for paper cup disposal may not be particularly large, and the nearest one may be at a significant distance. It may then occur that the network is not used to its full capacity, leading to congestion at some nodes and low levels at some other nodes.  Also, when the collectors come to collect bins in an area, it is not an efficient collection if some bins are full and some are not. \\ 

A possible solution for this problem could be to provide different levels of reward to different bins, to facilitate a balanced utilisation of all of them, and thus mitigate the possibility that the most popular ones become full soon. However, varying rewards may cause variations in the deposit or higher values of the deposit, and thus create practical issues in the prices of cups of coffee. Accordingly, in this work, we provide an alternative solution where the reward is constant, and we use time-varying signalling systems to advertise the availability of smart recycling bins, with the final objective of maximising the user's expected reward.\\

\subsection{Optimal Choice of Monetary Rewards}
\label{Problem_2}
\begin{figure}[H]
    \centering
    \includegraphics[width=1\linewidth]{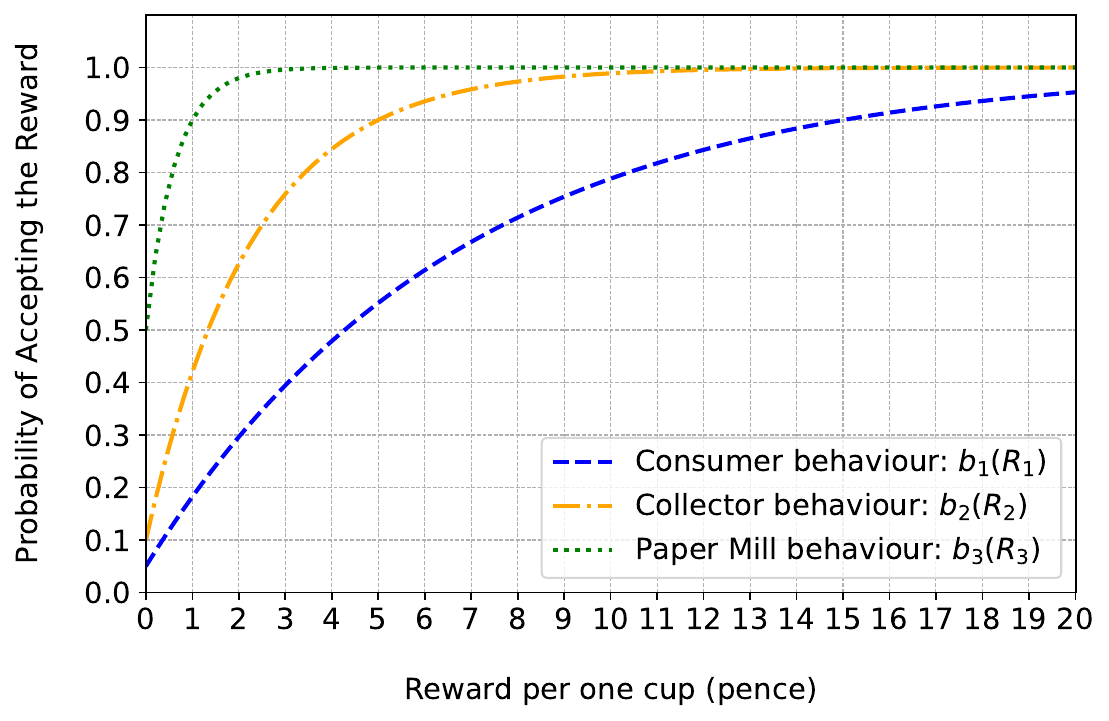}
    \caption{\small {Acceptance Probability of Offered Reward Across Different Agents in the Recycling Chain.}
        This graph illustrates the {probability of accepting a given reward} (pence per cup) for recycling, as modelled for three key agents in the recycling chain: {Consumers} (dashed blue line, $b_1(R_1)$), {Collectors} (dash-dotted orange line, $b_2(R_2)$), and the {Paper Mill} (dotted green line, $b_3(R_3)$). The curves represent the distinct behavioural responses of these agents to varying reward levels, where $R$ is the reward offered.
    }
    \label{fig:behaviours}
\end{figure}

It is known that increasing the value of a reward increases the probability of acceptance \cite{yu_effectiveness_2017}, e.g., that a customer will throw the paper cup in an appropriate bin to facilitate the recycling process onto the next level, rather than throwing it away in another -- nearer -- bin. Accordingly, the behaviour of participants can be modelled with a curve like the one depicted in Fig.~\ref{fig:behaviours}. However, each set of participants between two different layers is, in principle, different from the others, since the task is different. For example, as it is shown in Fig.~\ref{fig:behaviours}, the acceptable reward for the consumer to throw the paper cup in the smart recycling bins is much different from the acceptable reward for the waste collector to collect it. So an algorithm is required to calculate and distribute optimal rewards $R_i^\star$ to maximise throughput of the network. \\

In our work, we do so under the assumption that the initial deposit $D$ is given, and the behaviour of agent $i$ at level $i$, $b_i(R_i)$, is known and private to them. \\

\subsection{Optimal Design of the Deposit}
\label{problem_3}

The deposit plays a key role in the recycling process because the sum of all the rewards should be sufficient to allow the paper cup to move across the entire reverse logistics chain and reach the paper mill. In particular, the amount of the deposit has a direct influence on the recycling rate that can be achieved. In practice, choosing the correct deposit value is not straightforward. The right amount is not known in advance, because the system is complex and involves many factors that cannot be easily measured or predicted \cite{Complex_Haque_2000}. For this reason, the deposit should be adjusted using feedback rather than fixed at an arbitrary level. In addition, introducing a high deposit for a product may confuse or frustrate users, so it is important that the adjustment happens slowly. A gradual change gives people time to adapt and helps the system reach stable recycling behaviour. Later, if recycling becomes a regular habit for most users, the deposit can even be reduced without lowering the overall performance of the system.\\

In this paper, we design a PI-based feedback loop to control the value of the deposit at a slow time rate, to achieve a desired level of recycling.\\

\section{Congestion Management and Load Balancing}
\label{sec:load_balancing}

In this section, we address the load balancing problem described in Section \ref{Problem_1}, i.e., how to balance resources in one layer, in order to prevent congestion events from occurring. For instance, if all customers throw away their cups in the same bin, then the bin will soon be full and frustrated customers may rather throw away the paper cup in a non-recyclable bin.\\

 If we assume that requests to throw away a paper cup arrive in a sequential way, then the optimal allocation is to recommend each time the \textit{emptiest} bin. However, such a solution would require all bins to exchange information among themselves --- or to communicate their own level of paper cups to a central hub, in order to identify the \textit{emptiest} bin. Conversely, in this paper, we wish to solve the same problem in a decentralised fashion to mitigate the need to exchange such information. For this purpose, we rather design a decentralised algorithm that is inspired by earlier works on the allocation of electric vehicles to charging stations \cite{Moschella2021,  Stochastic} and parking spaces \cite{Delay}. In particular, we assume that bins broadcast an availability signal to customers more frequently when there is more residual space. Accordingly, when a customer needs to throw away a paper cup, they will select the bin that first sends out an availability signal, and more likely, this will be the most convenient bin. In particular, each bin uses the algorithm \ref{alg:poisson} to decide when to broadcast the availability signal.

\begin{algorithm}[H]
\caption{Run locally in bin $j$ to broadcast availability signal}
\begin{algorithmic}[1]
\State $r_j \gets$ residual space in bin $j$
\State $\lambda_j \gets F(r_j)$ \Comment{$F:\mathbb{R}_{\ge 0}\!\to\!\mathbb{R}_{\ge 0}$ is increasing}
\If{$\lambda_j > 0$} %\State \textbf{return} 
%\Else
\State $u \gets \mathrm{Uniform}(0,1)$ 
\State $t_w \gets -\ln(u)/\lambda_j$
\State \textbf{wait} $t_w$ seconds
\State \textbf{broadcast} availability signal
\EndIf
\State \textbf{go to} 1 \Comment{Recompute $\lambda_j$ each cycle}
\end{algorithmic}
\label{alg:poisson}
\end{algorithm}

In Algorithm~\ref{alg:poisson}, $r_j$ denotes the residual free space in bin $j$, and $F(r_j)$ is an increasing function mapping the residual capacity to a broadcast rate $\lambda_j = F(r_j)$ (differently from the algorithm in~\cite{Stochastic} where broadcasting probabilities had been considered instead). The variable $u$ is a random number selected from a uniform distribution in $(0,1)$, used to generate an exponential waiting time $t_w = -\ln(u)/\lambda_j$. After waiting $t_w$ seconds, the bin broadcasts its availability signal and then repeats the process with the updated value of $r_j$. Under Algorithm~\ref{alg:poisson}, an arbitrary bin $j$ sends signals at average rate $\lambda_j$ \cite{kelin_1984_Poisson}. Since $\lambda_j$ increases with $r_j$, bins with more (less) free space advertise more (less) frequently.\\

Then, the advantage of exploiting Poisson rates is that the probability that the $j$'th bin will broadcast the availability first among all bins follows the so-called standard competing exponential equation (or exponential race)~\cite{balakrishnan1995exponential}:

\begin{equation}
\Pr(\text{assign to } j)
= \frac{\lambda_j}{\sum_{k=1}^{n} \lambda_k}.
\label{eq:pr_assign}
\end{equation}

In our context, Equation (\ref{eq:pr_assign}) defines the stochastic rule by which each new request (e.g., a consumer) is assigned to a specific bin.\\

\begin{remark} So far, we have assumed that paper cups are assigned to bins depending solely on the fill level of the bin, to prevent congestion events from occurring (i.e., one bin is quickly full while the other bins are relatively empty). However, it is also important that distance is taken into account, so that when multiple feasible bins exist, bins that are excessively far away are not recommended. For this purpose, it is possible to consider $r_j$ to be a weighted combination of both the fill level of the $j$'th bin and also of the distance of the bin, so that if two bins have a similar level of space, then the closest one is recommended. This operation can be done similarly to what had been proposed in paper \cite{Moschella2021}.
\end{remark}

\subsection{Choice of Function $F(r_j)$}

In principle, any increasing function $F:\mathbb{R}_{\ge 0}\!\to\!\mathbb{R}_{\ge 0}$ is admissible.  
While~\cite{Stochastic} used a function roughly equivalent to $F(r_j)=10^{r_j}$, here we choose a simpler tunable power law:
\begin{equation}
F(r_j)=r_j^{a}, \quad a>0.
\end{equation}
Substituting this into~\eqref{eq:pr_assign} yields
\begin{equation}
q_j(a)
= \Pr(\text{assign to } j)
= \frac{r_j^{a}}{\sum_{k=1}^{n} r_k^{a}}.
\label{eq:q}
\end{equation}

Here, $a$ is a tunable parameter controlling the level of determinism in the decentralised assignment.  
While in general assignments favour bins with larger values of $r_j$, this behaviour is amplified for larger values of $a$, as we shall show in the next section.

\subsection{Convergence to the Deterministic Algorithm}

We now show that as $a \to \infty$, the stochastic decentralised policy~\eqref{eq:q} converges to the centralised deterministic policy, in which each new request is assigned to the emptiest bin, without requiring the exchange of information among the bins.

Let $r_{\max}$ denote the largest residual space available in the set of bins, i.e., $r_{\max} = \max_j r_j$ and let us define the index set
\[
Q = \{ j : r_j = r_{\max} \}.
\]
The set may contain more than one element if more than one bin has the same largest availability of space.
Then, for any $j$,
\[
\lim_{a \to \infty} q_j(a)
= \begin{cases}
1 / |Q|, & j \in Q, \\[4pt]
0, & j \notin Q.
\end{cases}
\]

Hence, as $a \to \infty$, all the probability mass concentrates on the bins with maximum residual capacity, i.e., emptiest ones. If only one bin attains $r_{\max}$, i.e., $|Q|=1$, then $\lim_{a\to\infty} q_j(a)=1$ for that bin, and every incoming request is assigned to it — exactly reproducing the deterministic rule “\textit{assign to the emptiest bin}”.\\

When multiple bins share the same maximum residual capacity, the stochastic rule distributes arrivals uniformly among them, which corresponds to a fair tie-breaking rule in the deterministic setting.  
Thus, in the limit $a \to \infty$, the decentralised algorithm behaves identically to the centralised deterministic algorithm under the assumptions that:
\begin{itemize}
    \item requests arrive sequentially (one by one), and
    \item there is no delay between assignment and action.
\end{itemize}
These conditions ensure that after each assignment, residual capacities are updated instantaneously, preserving synchrony between the decentralised stochastic decision process and the deterministic centralised policy.

\subsection{Simulation}

\subsubsection{Simulation Assumptions}
\begin{figure}[H]
  \centering
  \begin{subfigure}[b]{0.49\textwidth}
    \includegraphics[width=\textwidth]{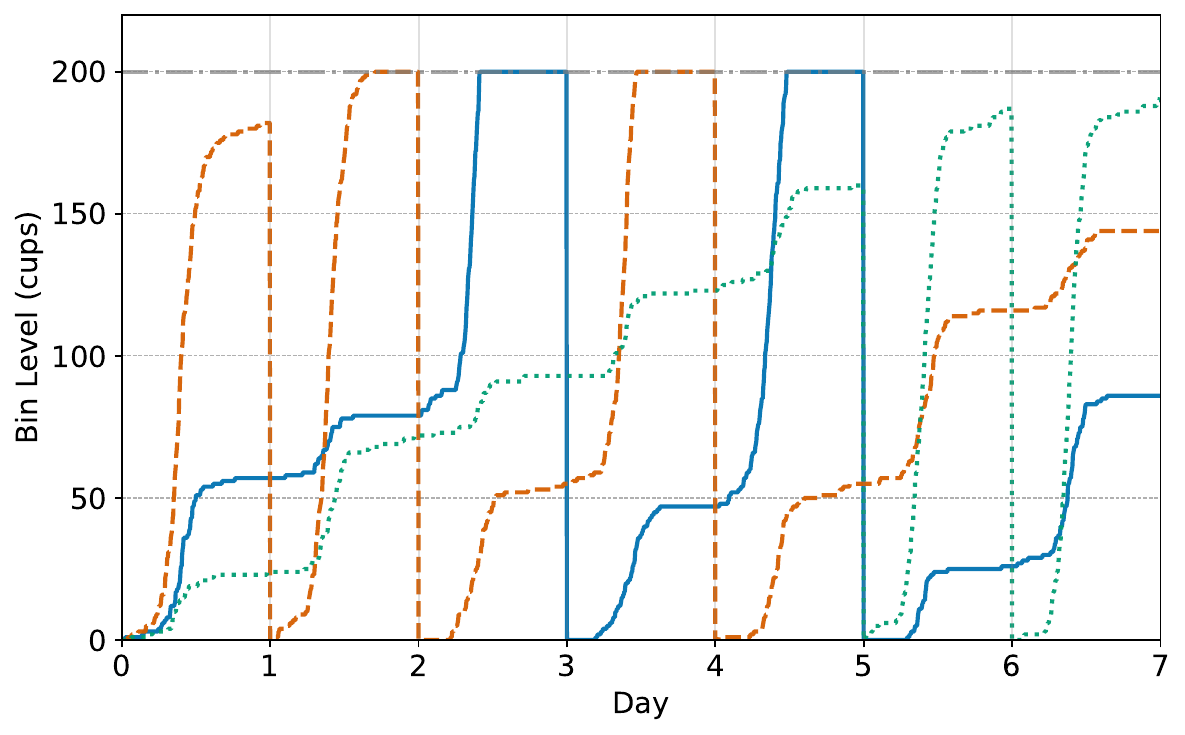}
    \caption{Unsupervised allocation}
    \label{fig:a}
  \end{subfigure}
  \hfill
  \begin{subfigure}[b]{0.49\textwidth}
    \includegraphics[width=\textwidth]{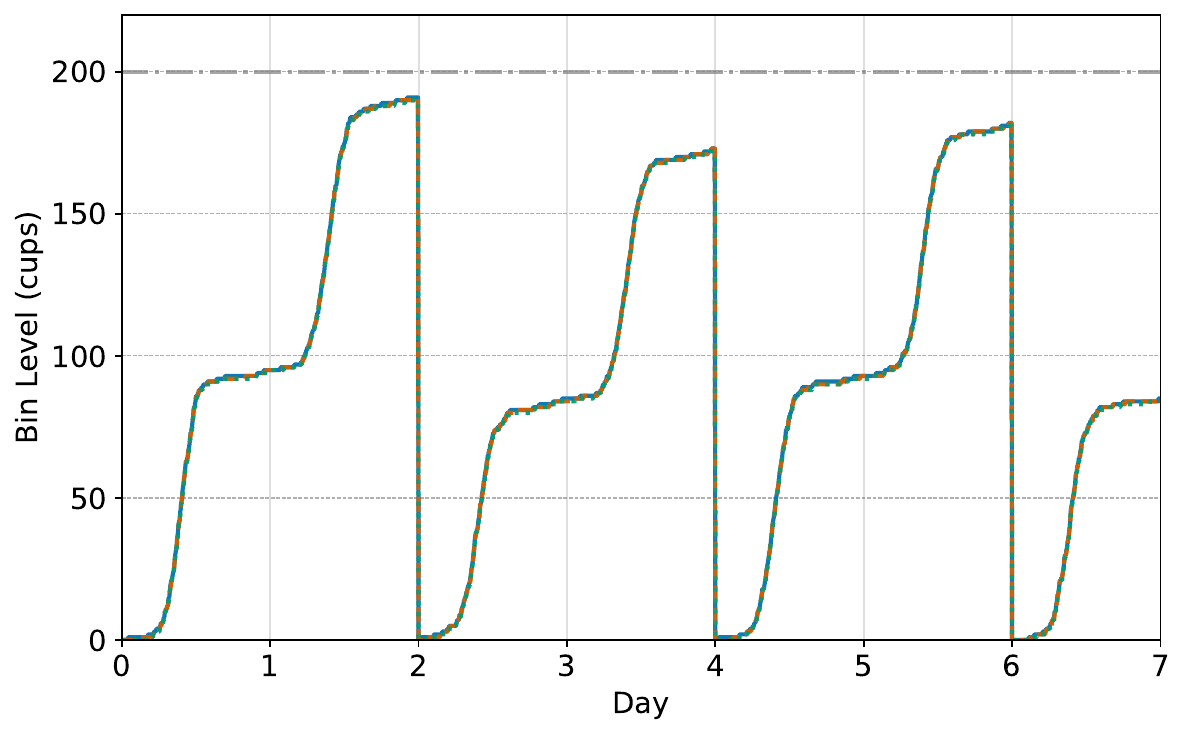}
    \caption{Optimal centralised allocation}
    \label{fig:b}
  \end{subfigure}

  \vspace{0.5cm}

  \begin{subfigure}[b]{0.49\textwidth}
    \includegraphics[width=\textwidth]{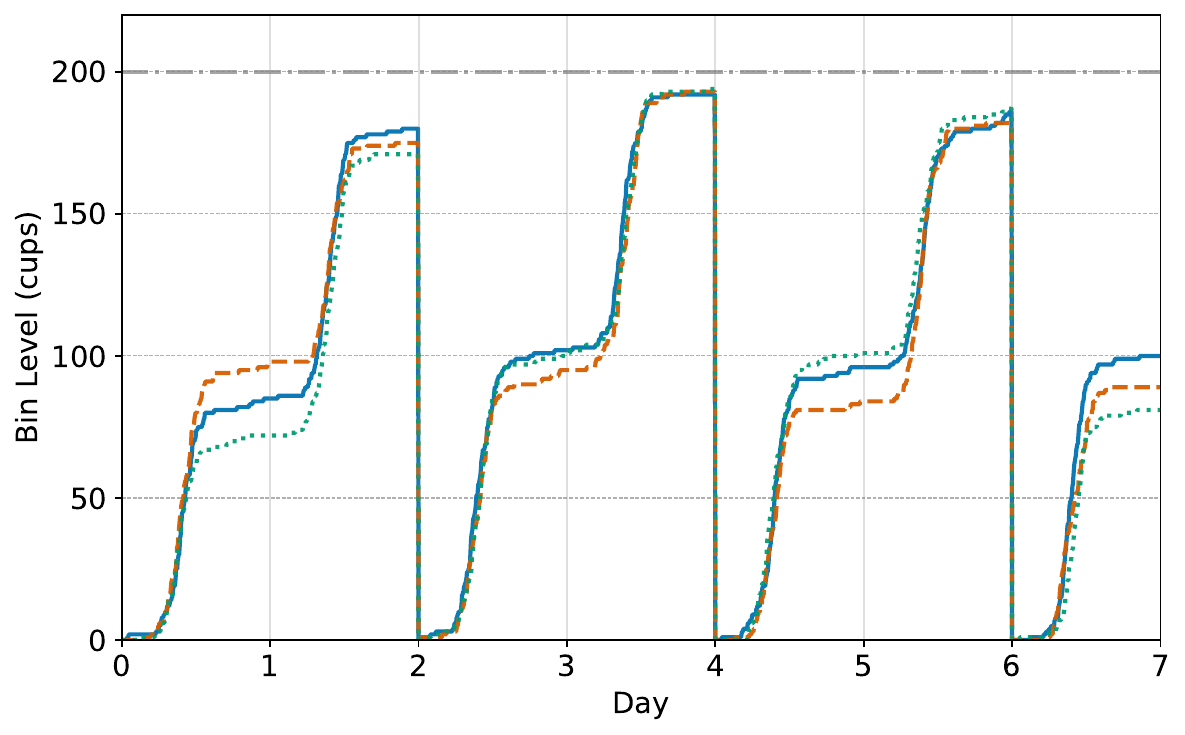}
    \caption{Decentralised algorithm with $a=1$}
    \label{fig:c}
  \end{subfigure}
  \hfill
  \begin{subfigure}[b]{0.49\textwidth}
    \includegraphics[width=\textwidth]{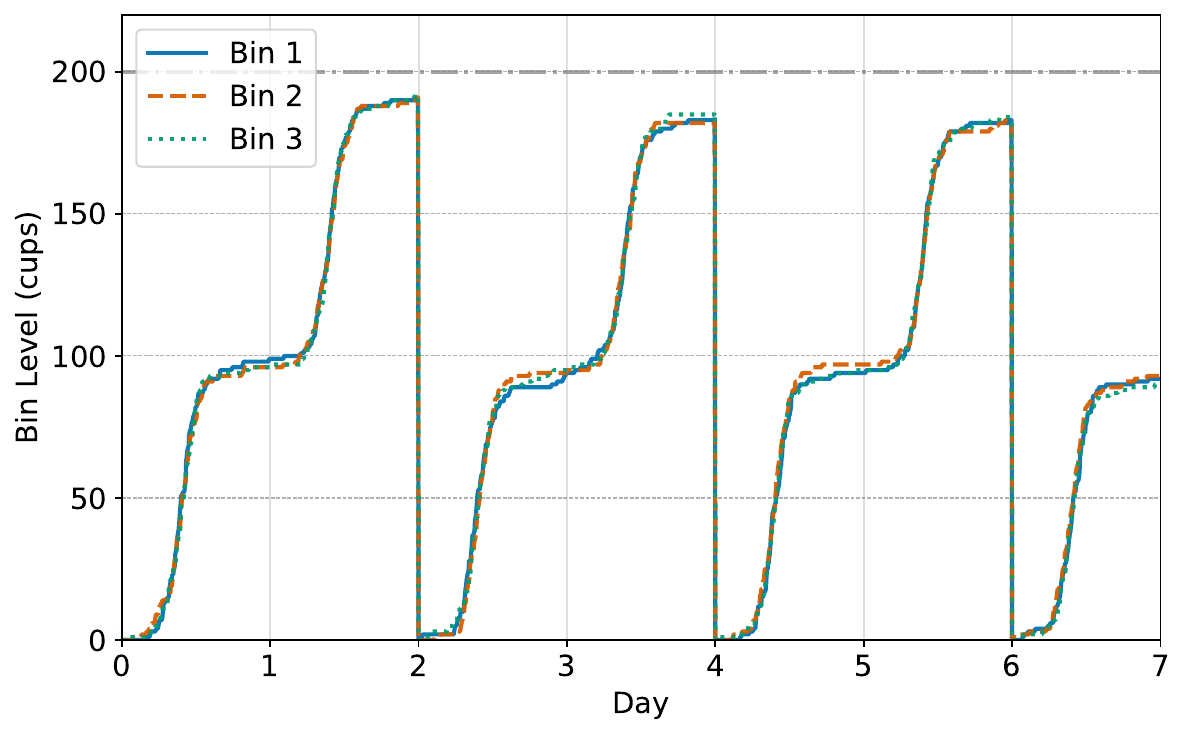}
    \caption{Decentralised algorithm with $a=8$}
    \label{fig:d}
  \end{subfigure}

\caption{\small Simulation results for a three-bin recycling system under four allocation strategies: (a) unsupervised allocation, showing an unbalanced solution with congestion and overflow; (b) optimal centralised deterministic assignment, producing perfectly balanced bins; (c) decentralised stochastic algorithm with $a=1$, achieving near-optimal balance; and (d) decentralised algorithm with $a=8$, showing almost identical behaviour to the centralised case. Simulations run at one-minute resolution for seven days, with bins emptied overnight when exceeding 75\% capacity.}
  \label{fig:matrix}
\end{figure}

We now simulate the behaviour of the proposed allocation strategy in a simplified scenario, to showcase the advantage with respect to plain approaches where a customer throws the paper cup in an unsupervised fashion (e.g., to the closest one). For this purpose, we consider for simplicity only three bins, where about 270 consumers per day need to throw away a paper cup according to a non-homogeneous Poisson process with higher intensity during morning hours, and a peak around 8~AM.  We assume that the bins have a capacity of 200 paper cups, and that a bin is emptied during the night if its level in the evening exceeds 150 paper cups (i.e., 75\% of the capacity). We run a simulation over a one-week period, comparing four different allocation strategies: (a) unsupervised allocation, where users choose bins according to personal preferences, e.g., the closest one (see Fig.~\ref{fig:a}). We assume that the popularity of one bin over the other may change according to weekly patterns (e.g., weekdays vs. weekends); (b) a centralised deterministic assignment, where each arrival is directed to the emptiest bin (see Fig.~\ref{fig:b}). We remind that this is the optimal assignment in the considered sequential framework. (c) our decentralised algorithm with $a = 1$ (which corresponds to a probabilistic proportional allocation, see Fig.~\ref{fig:c}); and (d) our decentralised algorithm with $a = 8$ (as we remind that for values of $a$ that tend to infinity the optimal deterministic behaviour, see Fig.~\ref{fig:d}).  \newline

\subsubsection{Discussion on the simulation results}

Simulation results are displayed in Fig.~\ref{fig:a}–\ref{fig:d}. In particular, Fig.~\ref{fig:a}  demonstrates that without any supervision, the level of the bins is unbalanced, and in this example, about 13\% of cups are not correctly recycled because the bin is full. In addition, the unsupervised solution is not only wasteful but also makes the collection procedure chaotic and inefficient. The centralised deterministic policy eliminates waste and perfectly balances loads. However, all bins should exchange information with each other in real-time -- or should communicate to a central entity -- in order to identify the emptiest bin.  Our proposed decentralised allocation algorithm produces results that are very well balanced already with $a=1$, and results that with $a=8$ are almost indistinguishable from those of the centralised solution, confirming that the proposed stochastic mechanism converges to the deterministic solution as the exponent $a$ increases. At the same time, the bins do not have to communicate with each other, and no central entity is required either. \\

\begin{remark}
Note that as the exponent $a$ increases, $t_w$ (inter-arrival times in signal broadcasting) increases. A large value of $t_w$ implies that the customer has to wait for a long time before the allocated bin is communicated, and this behaviour is undesirable to prevent the customer from losing interest in waiting and just throwing away the paper cup in an inappropriate bin. Besides, simulation results (Fig. \ref{fig:d}) and mathematical analysis show that centralised and decentralised algorithms provide comparable outcomes for small values of $a$ as well. \\
\end{remark}

\begin{remark}
    Although our load balancing algorithm reduces congestion and balances the loads among the nodes, from the users' point of view, a key modelling choice is whether incentives are defined as a guaranteed reward at each individual node or in terms of the users' expected reward. We adopt the latter perspective. Rather than promising a fixed reward at every node -- which may not always be delivered due to congestion or lack of capacity -- we assume that each agent decides based on the expected reward it receives over time, given the stochastic dynamics of the system. This viewpoint is closer to how participants experience the scheme in practice: rational users will follow the proposed load balancing policy so as to maximise their expected reward. By reducing congestion across nodes, the system increases this expected reward and thus strengthens participation incentives. 
\end{remark}

\section{Reward pricing with AIMD }
\label{sec:AIMD}

In this section, we address the problem of optimally computing the monetary rewards along the recycling stage, as anticipated in Section~\ref{Problem_2}. When a fixed monetary amount must be divided among multiple participants, classical fairness criteria such as the Shapley value are often used \cite{Shapley,Cooperating_Participants,Shared_Network_Infrastructures}. Alternatively, the deposit can be interpreted as a shared resource, making the assignment of rewards a resource-allocation problem within the network utility optimisation framework \cite{Tutorial,Cent_NUO}. Decentralised formulations based on utility optimisation are also available \cite{Decen_NUO}, and among these, {AIMD} algorithms are particularly attractive due to their low communication requirements and scalability \cite{Bob_AIMD,Ferraro01092024}. The objective here is to maximise the \textit{throughput}, i.e., the percentage of paper cups that are recycled at the end of the process. The problem can be mathematically formulated as in~(\ref{eq:optimization_problem}):

\begin{equation}
\begin{array}{ll}
\max\limits_{R_1, \ldots, R_L} & \prod\limits_{i=1}^{L} b_i(R_i) \\[6pt]
\text{s.t.} & 
\begin{cases}
\sum\limits_{i=1}^{L} R_i \leq D, \\
 R_i \geq 0, \forall i.
\end{cases}
\end{array}
\label{eq:optimization_problem}
\end{equation}

In the optimisation problem (\ref{eq:optimization_problem}), $b_i(R_i)$ is the function that describes the behaviour of agents in layer $i$. In particular, the functions relate the probability of appropriately taking the correct recycling action to the value of the monetary reward $R_i$ (see Figure \ref{fig:behaviours} for an example). In the optimisation problem (\ref{eq:optimization_problem}), $L$ denotes the number of layers in the recycling process (e.g., here $L = 3$ ).\\

\subsection{Centralised solution}
\label{Centralised_Solution}

Problem (\ref{eq:optimization_problem}) may be easily solved in a centralised way if all the behaviour functions $b_i({\cdot})$ were known in advance to a centralised entity. For instance, using a Lagrangian approach \cite{Boyd2004ConvexO}, we may rewrite the optimisation problem as:

\begin{equation}
    \mathcal{L}(R_1, \ldots, R_L, \lambda) = \prod_{i=1}^{L}b_i(R_i) - \lambda \left(-D + \sum_{i=1}^{L} R_i \right),
    \label{Lagrangian_Solution}
\end{equation}

and by taking partial derivatives, we obtain

\begin{equation}
    \frac{\partial \mathcal{L}}{\partial R_i} = b_i'(R_i) \cdot \prod_{\substack{j=1 \\ j \neq i}}^{L}b_j(R_j) - \lambda = 0 \quad \Rightarrow \quad b_i'(R_i) \cdot \prod_{\substack{j=1 \\ j \neq i}}^{L}b_j(R_j) = \lambda
    \label{eq:partial_lagrange_reward}
\end{equation}
\begin{equation}
    \frac{\partial \mathcal{L}}{\partial \lambda} = D - \sum_{i=1}^{L} R_i = 0
    \label{eq:partial_lagrange_lambda} \quad \Rightarrow \quad \sum_{i=1}^{L} R_i = D
\end{equation}
Equations (\ref{eq:partial_lagrange_reward}) and (\ref{eq:partial_lagrange_lambda}) may be used to form a system of equations with $L+1$ unknowns and $L+1$ equations, which may be solved either analytically or numerically, depending on the nature of the behaviour functions $b_i(R_i)$. Also, note that in general, from Equation (\ref{eq:partial_lagrange_reward}), it is possible to observe that the optimal solution is achieved when consensus occurs among the ratios

\begin{equation}
    \frac{b_1(R_1)}{b'_1(R_1)} = \frac{b_2(R_2)}{b'_2(R_2)} = \ldots = \frac{b_i(R_i)}{b'_i(R_i)} = \ldots = \frac{b_L(R_L)}{b'_L(R_L)}.
    \label{eq:consensus}
\end{equation}

\subsection{Decentralised solution}
\label{Decentralised_Solution}

A centralised solution to the optimisation problem (\ref{eq:optimization_problem}) was described in Section \ref{Centralised_Solution}. However, since usually different actors are involved in the recycling chain, it may easily occur that not all of them are willing to disclose such private functions, and the interest is in solving the optimisation problem in a decentralised manner. For this purpose, we solve it using the classic decentralised AIMD algorithm \cite{Bob_Book}, which has been widely adopted in many frameworks, e.g., most notably, to optimally share bandwidth in Internet congestion in a decentralised manner \cite{Chiu}. According to the algorithm, agents (here, the different layers in the recycling process) linearly increase their request for reward at each step (Additive Increase (AI) stage), until the sum of all the rewards exceeds the value of the deposit (i.e., the first constraint in (\ref{eq:optimization_problem})). In the classic synchronised AIMD framework, when the constraint is exceeded, then agents perform a Multiplicative Decrease (MD) step to reduce their utilisation of the resource and recover a feasible solution. In the unsynchronised version of AIMD, only a subset of the agents perform the MD step, while the others keep increasing their reward. Such an unsynchronised AIMD is particularly convenient in optimal resource allocation problems \cite{Bob_Book}-- as the one of interest here, where we wish to optimally allocate rewards within the available deposit. Accordingly, as described in line 8 of the Algorithm \ref{alg:AIMD}, only a subset of rewards is decreased, according to a probability that depends on their behaviour functions $b_i({\cdot})$. In this way, the values of the rewards converge to the optimal solution of the optimisation problem (\ref{eq:optimization_problem}), without requiring communication among the layers. The only required communication occurs when a central agent signals the congestion event, i.e., that the sum of rewards is exceeding the deposit. The AIMD algorithm may be described as in Algorithm \ref{alg:AIMD}.

\begin{algorithm}[H]
\caption{{\small AIMD-based algorithm run locally in layer $i$ to maximise the overall throughput of the network.}}
\begin{algorithmic}[1]
\Statex \textbf{Initialisation :} The layers set their initial rewards to arbitrary values, $R_i(1)$, and the parameter, $\Gamma>0$, is broadcast to all of them; 
\Statex $k=0$, $l=0$\\
\textbf{Input:} $D > 0$, $\alpha > 0$ , $\beta\in (0,1)$  
%\Statex \hspace{2.5cm} The parameter $\Gamma$ is broadcast;
\vspace{0.2cm}
\For{$l\leftarrow l+1$}
    \vspace{0.2cm}
    \If{$\sum_{i=1}^{L} R_i(l) \leq D$}
        \vspace{0.2cm}
        \State $R_i(l+1) = R_i(l) + \alpha$ \Comment{Additive Increase phase}
    \vspace{0.2cm}
    \Else \Comment{This is a capacity event}
            \vspace{0.2cm}
            \State $k\leftarrow k+1$
            \vspace{0.2cm}
            \State $R_{i,c}(k) \leftarrow R_i(l)$
            \vspace{0.2cm}
            \State $\bar{R_i}(k) \leftarrow \frac{1}{k} \sum_{j=1} ^k R_{i,c}(j)$
            \vspace{0.2cm}
            \State $R_i(l+1) = \begin{cases}
            \beta \cdot R_i(l) & \text{with probability } \pi_{i}(\bar{R_i(k)}) \quad \text{calculated by \ref{eq:lambda} } \\
           
            R_i(l)+\alpha  & \text{with probability } 1 - \pi_{i}(\bar{R_i}(k))
            \end{cases}$
    \vspace{0.2cm}
    \EndIf
\vspace{0.2cm}
\EndFor
\end{algorithmic}
\label{alg:AIMD}
\end{algorithm}

\textbf{Proposition 1:} In the unsynchronised AIMD algorithm, if the objective function is maximising the product of concave utility functions, then the probability of decrease for layer $i$ is calculated as:
\begin{equation}
    \pi_i(\bar{R_i}) = \Gamma \frac{b_i(\bar{R_i}(k))}{\bar{R_i}(k)  b'_i(\bar{R_i}(k))},
    \label{eq:lambda}
\end{equation}
where $\bar{R_i}(k)$ is the long-term average reward, see line 7 of Algorithm \ref{alg:AIMD}. Details about the derivation of Equation (\ref{eq:lambda}) are provided in subsection \ref{sec:proof}.\newline

By running Algorithm \ref{alg:AIMD} locally in each layer, the rewards $R_i$ converge to their optimal values $R_i^{*}$, in a privacy-preserving (decentralised) way. It means that not only do different agents of the recycling network not know the behaviour function of the others, but also the system operator does not require knowledge of the functions.\\

\begin{remark}
The behaviour functions, \( b_i(R_i) \), are private to each agent. If an agent misrepresents or misestimates its function, a single stage of the recycling process may be over-rewarded or under-rewarded. To prevent such discrepancies from occurring, we propose that the central authority (i.e., the system operator) has access to approximate upper and lower thresholds for each agent’s behaviour at the optimal reward, i.e., \( \underline{B}_i < b_i(R_i^*) < \overline{B}_i \). Or, in other words, the system operator should have an indicative range of success of a given stage in the recycling process (e.g., 70 to 90 \% of the paper cups should be thrown in the appropriate bins). If one of the targets is missed, then the operator may recommend adjustments and advise the corresponding agent \( i \) to increase or decrease the $b_i(R_i)$. Importantly, the exact form of the behaviour functions and the precise optimal rewards remain private; the central authority only relies on shared estimates of the behaviour range.
\end{remark}

\subsubsection{AIMD formulation of the reward computation}
\label{sec:proof}

The original formulation of the unsynchronised AIMD network utility maximisation \cite{Bob_Book}, which is used in many practical applications, see for instance \cite{7801051}, is as follows:

\begin{equation}
\begin{array}{ll}
\max\limits_{R_1, \ldots, R_L} & \sum\limits_{i=1}^{L} g_i(R_i) \\[6pt]
\text{s.t.} & 
\begin{cases}
\sum\limits_{i=1}^{L} R_i \leq D, \\
 R_i \geq 0, \forall i.
\end{cases}
\end{array}
\label{eq:AIMD_network_utility_optimisation}
\end{equation}

where $g_i(R_i)$ are concave functions. In this set-up, if only a subset of the agents perform the multiplicative decrease step of Algorithm \ref{alg:AIMD} according to probability

\begin{equation}
    \Gamma \frac{1}{\bar{R_i}(k) \, g'_i(\bar{R_i}(k))}
    \label{eq:AIMD_pi},
\end{equation}
then the long-term averages of the rewards converge to the optimal values.

We can turn our optimisation problem (\ref{eq:optimization_problem}) into the standard AIMD form (\ref{eq:AIMD_network_utility_optimisation}) by taking the natural log. Indeed, maximizing the product of \( b_i \) is equivalent to maximizing its logarithm:
\begin{equation*}
\max \prod_{i=1}^{L} b_i(R_i) \quad \equiv \quad \max \ln\left( \prod_{i=1}^{L} b_i(R_i) \right).
\end{equation*}

Using the logarithmic identity \( \ln\left( \prod_{i} b_i(\cdot) \right) = \sum_{i} \ln(b_i(\cdot)) \), this is equivalent to:
\begin{equation*}
\max \sum_{i=1}^{L} \ln(b_i(R_i)).
\end{equation*}

Accordingly, if we define \(g_i(R_i) \coloneqq \ln(b_i(R_i))   \), we obtain the standard form of AIMD for our optimisation problem, under the condition that \( \ln(b_i(R_i)) \) is concave. Embedding our definition of $g_i(R_i)$ into Equation (\ref{eq:AIMD_pi}), we obtain the previously introduced Equation (\ref{eq:lambda}) for the unsynchronised multiplicative step.

\subsection{Simulations}

\subsubsection{Data for Paper Cups Case study}
Research from the House of Commons Environmental Audit Committee indicates that the UK uses around 2.5 billion disposable coffee cups annually\footnote{\href{https://nationwidewasteservices.co.uk/the-environmental-impact-of-disposable-coffee-cups-in-the-uk/}{nationwidewasteservices.co.uk}}. Using London population data (13\% of the UK), we can estimate that this number for London is 325 million annually, or 890 thousand per day. A study by the University of Kent estimates that only 0.25\% (one in four hundred) of these paper cups in the UK are recycled\footnote{\href{https://www.kent.ac.uk/news/environment/32807/storm-in-a-coffee-cup-how-the-disposable-cup-problem-is-reaching-boiling-point?}{kent.ac.uk/news/environment/}}. {This extremely low recovery rate is attributed to three main barriers: (i) limited consumer awareness about why and how cups should be returned; (ii) insufficient public recycling infrastructure, including a shortage of dedicated cup-collection bins; and (iii) material and processing constraints, as most cups contain polymer linings that are expensive for standard recycling or kerbside collection.} Accordingly, this is a relevant problem with a significant margin of improvement.

\subsubsection{Simulation Assumptions}
We consider now the recycling network investigated so far, with the three agents of consumers, collectors, and paper mills. We choose concave utility functions to guarantee convergence of the AIMD algorithm, of the form:
\begin{equation}
    b_i(R_i; p_{0}, x_{90}) = 1 - (1-p_{0})\exp\left(-\frac{\ln\left(\dfrac{1-p_{0}}{0.1}\right)}{x_{90}} \times R_i\right)
    \label{eq:behaviour}.
\end{equation}
In functions (\ref{eq:behaviour}), $p_0$ is the base probability, i.e., the probability that the recycling action is performed correctly even without a reward, which is a parameter that can be easily estimated (e.g., how many paper cups are thrown in the correct bin even without economic reward). Parameter $x_{90}$ denotes the reward price that is required to achieve a probability of recycling equal to $90\%$. This second parameter is harder to estimate, but can be reasonably estimated in an approximate fashion from available recycling data \cite{ZHOU_2020_DRS_Review}. While we consider this function for the sake of simplicity, any other continuous concave function, differentiable, increasing and bounded between 0 and 1, is acceptable as well. Also, the agent's behaviour curves can be extrapolated and fitted from measured data. Examples of such functions were provided in Figure \ref{fig:behaviours}, with parameters given in Table \ref{Table_behaviour_parameters}.

\begin{table}[h!]
\centering
\caption{\small Parameters of behaviour functions}
\label{Table_behaviour_parameters}
\begin{tabular}{|l|c|c|}
\hline
Recycling phase / Parameters & $p_0$ & $x_{90}$ \\ \hline
Consumers & $5\%$ & $0.15$ \pounds \\ \hline
Collectors & $10\%$ & $0.05$ \pounds \\ \hline
Paper mills & $50\%$ & $0.01$ \pounds \\ \hline
\end{tabular}
\end{table}

In Table \ref{Table_behaviour_parameters}, base probabilities $p_0$ for consumers, collectors, and paper mills are selected to be consistent with the measured data of the current recycling rate of $0.25\%$, without monetary rewards. For the selection of the prices for the $90\%$ acceptance (i.e., $x_{90}$), there are no published data for paper cup deposits. However, some metrics on beverage containers~\cite{reloop2024} show that return-to-retail schemes with deposits around \pounds0.15 achieve return rates close to 90\%, and similar numbers are also reported in the IOTA KupKrush scheme \footnote{\href{https://biotasphere.com/}{https://biotasphere.com/}}. {Consistent with international DRS data~\cite{reloop2024}, these observations indicate that higher deposit levels generally lead to higher return rates.} Our parameters and behaviour functions are based on such measured data. \\

\subsubsection{Discussion on the simulation results}
\begin{figure}[H]
    \centering
    \includegraphics[width=1\linewidth]{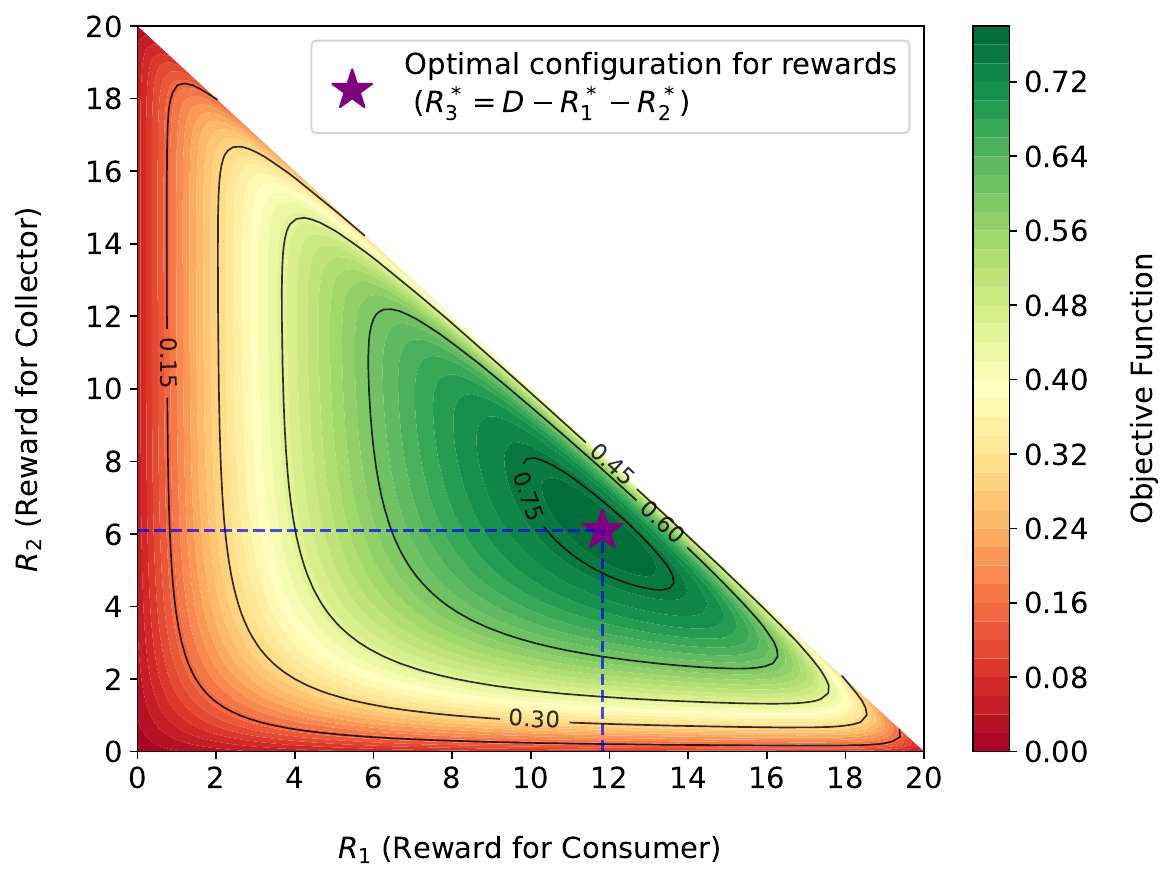}
    \caption{\small Overall recycling rate as a function of how the deposit is split into rewards for the different stages of the recycling process. Greener regions represent reward configurations that yield higher recycling rates, while redder regions correspond to poorer configurations. The star marks the optimal reward configuration that maximises the recycling rate, obtained from a centralised solution for a fixed deposit of $D=20$. At this point, the optimal rewards are $R_1^{*}=12$ for consumers and $R_2^{*}=6$ for collectors; by constraint~\ref{eq:R_i<D}, the remaining reward is $R_3^{*}=2$ pence per cup for the paper mill.}
    \label{fig:cen}
\end{figure}

Based on the previously defined behaviour functions, Figure~\ref{fig:cen} shows how the overall recycling rate varies as the deposit is allocated across the different stages of the recycling process, with values ranging from as low as $0.5\%$ to approximately $77\%$. Greener regions indicate reward configurations that yield higher recycling rates, while redder regions represent poorer configurations. According to the centralised optimisation for a fixed deposit of $D=20$, the optimal reward split is approximately $R_1^{*}=12$ pence for consumers and $R_2^{*}=6$ pence for collectors; by constraint~\ref{eq:R_i<D}, this implies $R_3^{*}=2$ pence per cup for paper mills. The star in the figure marks this optimal configuration, which maximises the overall recycling rate under the assumed behavioural model.\\

\begin{figure}[H]
    \centering
    \includegraphics[width=1\linewidth]{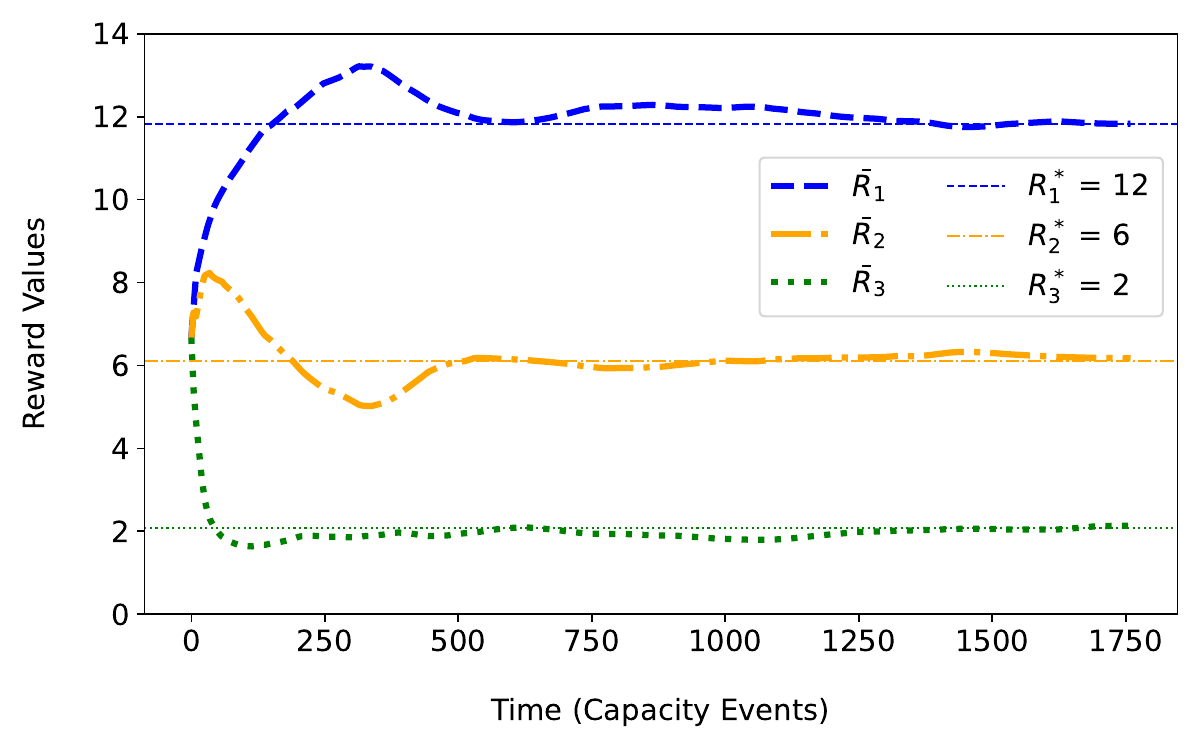}
    \caption{\small Evolution of reward values ($\bar{R}_i$) from a same point (or any arbitrary point) to their optimal values, through the AIMD-based algorithm \ref{alg:AIMD}.} 
    \label{fig:AIMD}
\end{figure}

Figure~\ref{fig:AIMD} illustrates how the AIMD-based algorithm \ref{alg:AIMD} drives the reward values $\bar{R}_1$, $\bar{R}_2$, and $\bar{R}_3$ toward the optimal configuration identified in Figure~\ref{fig:cen} (namely $12$, $6$, and $2$ pence per cup). Starting from arbitrary initial conditions, each layer updates its reward using only local information and without disclosing its behaviour function. The trajectories show clear convergence to the optimal values obtained from the centralised solution, demonstrating that the proposed decentralised algorithm can reliably reproduce the same optimal split of the deposit across the three stages of the recycling process. This confirms that the AIMD mechanism enables the actors to self-adjust their rewards in a fully decentralised manner while still achieving the maximum recycling rate.\\

\begin{figure}[H]
    \centering
        \includegraphics[width=1.0\linewidth]{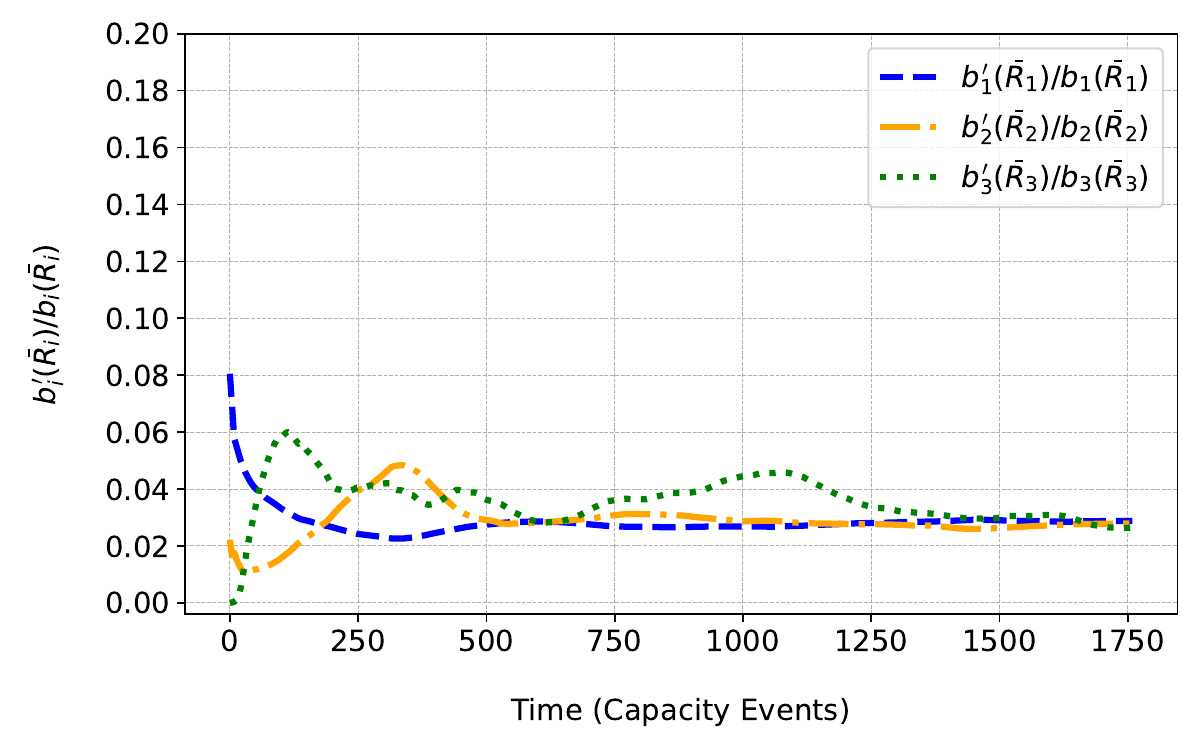}
        \caption{\small Evolution of $b_i'(\bar{R}_i)/b_i(\bar{R}_i)$ to consensus. This figure satisfies Eq. \ref{eq:consensus}. So, it validates the convergence of our proposed AIMD-based Algorithm to the optimal point.}
        \label{fig:consensus2}
\end{figure}

Figure~\ref{fig:consensus2} shows the evolution of the ratios $b_i'(\bar{R}_i)/b_i(\bar{R}_i)$ over time, demonstrating that for all layers it converges to the same value. This consensus property is precisely the optimality condition expressed in Eq.~\ref{eq:consensus}, indicating that the decentralised AIMD dynamics not only steer the reward values toward the optimal configuration (as shown in Figure~\ref{fig:AIMD}) but also satisfy the theoretical requirement for optimality. 

\section{Deposit Feedback Loop}
\label{sec:PID}

\begin{figure}[H]
    \centering
    \includegraphics[width=1\linewidth]{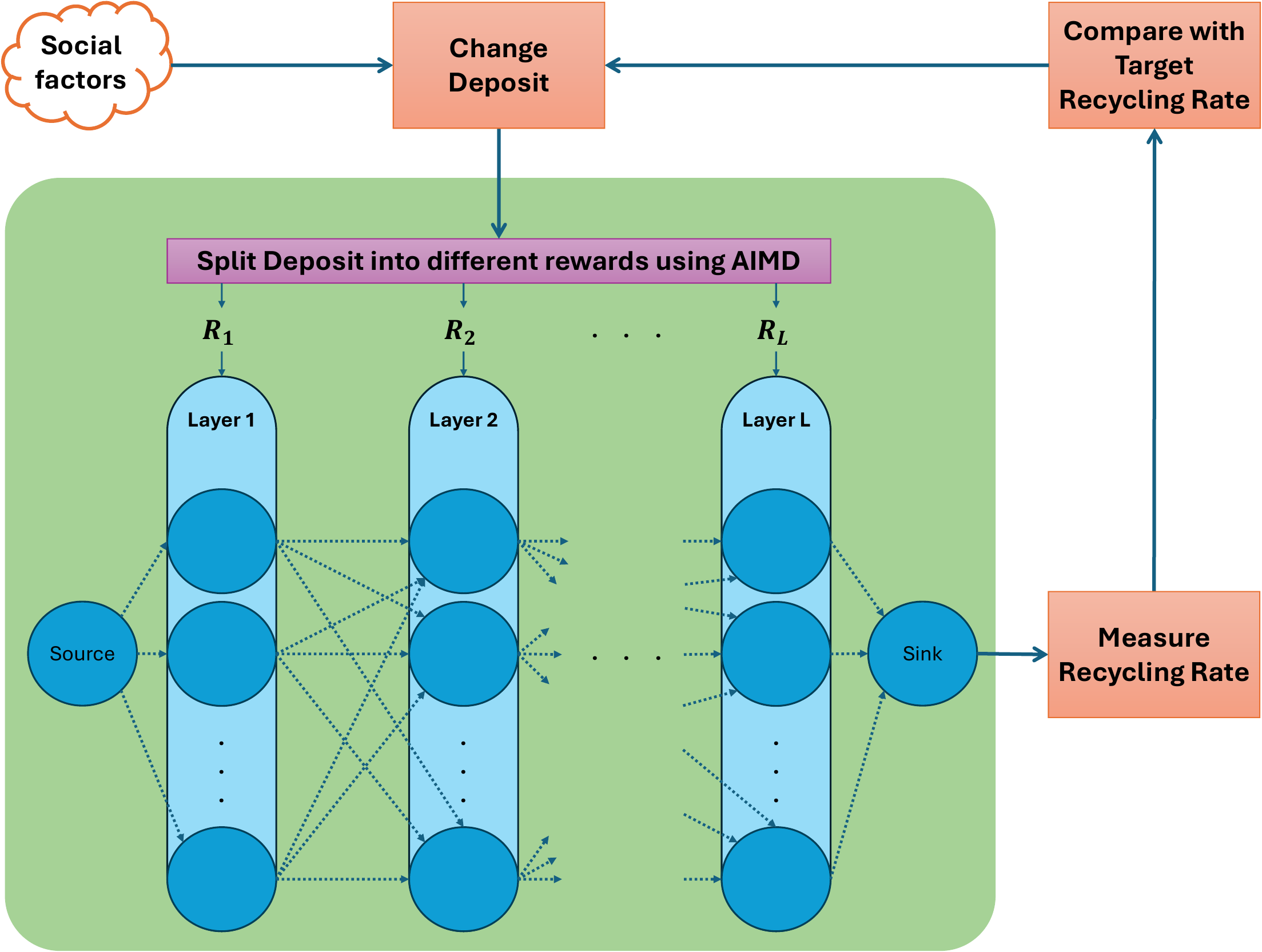}
    \caption{\small A closed-loop feedback control system over the network for controlling deposit values, to achieve a target recycling rate. The control system adjusts deposit levels based on measured recycling rate and social factors. An AIMD algorithm splits the new deposit into layer-specific rewards ($R_1, R_2, \dots, R_L$).}
    \label{fig:Control}
\end{figure}

In this section, we address the problem of adjusting the deposit value described in Section~\ref{problem_3}, namely how to tune the overall deposit so that the recycling rate reaches a desired target level. While several studies determine deposit levels through static or infrequently updated schemes---such as profit-maximisation formulations \cite{Taebok_Kim_2024}, fixed-policy analyses \cite{Gorgun2021, Butler2025, spataru2024short}, or economic evaluations based on break-even or willingness-to-pay considerations \cite{Schneider2021Economic, Linderhof2019, picuno2025potential}---these approaches do not provide mechanisms for real-time adjustment. After guiding consumers to the best bin and allocating the deposit across layer-specific rewards, if the overall throughput (e.g.\ recycling rate) remains below the target, the system can respond by gradually adjusting the initial deposit upward. Motivated by the use of feedback-control techniques in dynamic pricing \cite{Hawkins2014PID,Badi2023ForeignTrade}, and by related PID-based monetary-policy models \cite{Shepherd2019PID}, we close a feedback loop over the network to adjust the deposit value (see Fig.~\ref{fig:Control}). \\

\begin{remark}
{It is important to clarify that the dynamic deposit designed in this paper is not a fast or high-frequency pricing mechanism. In contrast to dynamic pricing schemes that change rapidly and may inconvenience users, our approach adjusts the deposit slowly and in gradual steps. This avoids sudden jumps that could confuse or frustrate consumers. The deposit increases only when the recycling rate remains below the target, and it does so in small increments. Similarly, once recycling becomes a stable habit for most users, the feedback loop allows the deposit to decrease slowly without harming overall system performance. This slow and gradual behaviour is essential for maintaining user convenience and acceptance.}
\end{remark}

\subsection{Simulation for Deposit feedback loop}

\subsubsection{Simulation Assumptions}
We simulate the deposit controlled recycling network for disposable paper cups in London, where it is reported that approximately 900{,}000 paper cups are consumed daily. Our simulation consists of three stages: (i) consumers to smart bins (daily), (ii) bins to collection centres (every two days, based on results in Section~\ref{sec:load_balancing}), and (iii) collection centres to paper mills (every five days). In the simulation, cups are generated according to a Poisson process with a rate of 900 per day (to mimic the actual 900,000 per day), and each is represented by a digital twin containing key attributes such as birthday, waste day, and waste location, along with a digital wallet holding its refundable deposit. The total deposit is allocated across the three layers: smart bins, collection centres, and paper mills, using an AIMD algorithm that determines optimal layer-specific rewards ($R_1$, $R_2$, $R_3$). These rewards probabilistically influence agent behaviour, increasing the probability of correct actions (i.e., transferring cups through the chain until the final stage), thereby improving the overall recycling rate. At the end of each month, a PI controller updates the deposit value based on the deviation between the achieved and target recycling rates, forming a closed feedback loop that dynamically tunes incentives to sustain the desired system performance (see Fig.~\ref{fig:Control}). Social factors also constrain the change in deposit: since consumers pay the deposit, sharp increases in its value would not be tolerated. Therefore, in this simulation, the change in deposit is limited such that it cannot increase by more than a factor of two in a single adjustment. After each deposit adjustment, the AIMD algorithm operates to split the deposit into optimal layer-specific rewards.

\subsubsection{Discussion on the simulation results}
\begin{figure}[H]
    \centering
    \includegraphics[width=1\linewidth]{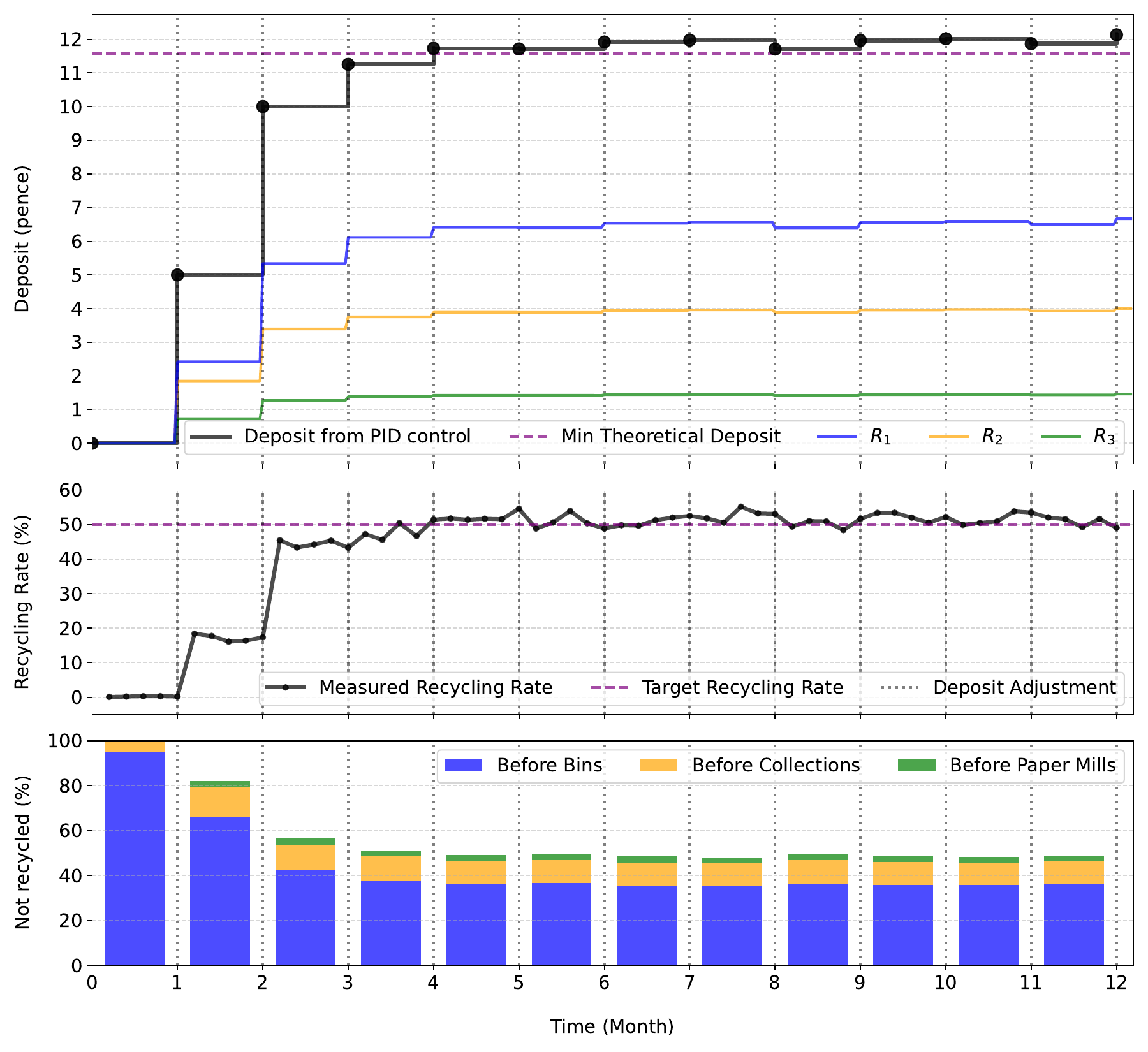}
    \caption{\small Simulation results for the deposit feedback loop. The upper plot shows the evolution of the deposit under PI control alongside the layer-specific rewards ($R_1$, $R_2$, $R_3$) computed via the AIMD algorithm. The middle plot shows the recycling rate converging toward its target. The lower plot shows, for each period and each stage, the percentage of cups that were not recycled.}

    \label{fig:PID}
\end{figure}

Fig.~\ref{fig:PID} presents part of the simulation results for the deposit feedback loop. As shown, the recycling rate increases and converges toward the target level as the PI controller dynamically adjusts the deposit value over time. The upper plot illustrates how the deposit evolves under PI control, together with the corresponding layer-specific optimal rewards ($R_1$, $R_2$, $R_3$) determined by the AIMD algorithm. The middle plot shows the recycling rate response, demonstrating the system’s stability and its ability to reach the desired target. The lower plot illustrates the percentage of the cups that were not recycled associated with each stage for each period, showing that the minimum total waste happens in an unequal combination of layer-specific wastes. These results confirm that the proposed feedback mechanism effectively regulates the deposit value, ensuring that the combined AIMD–PI control maintains system performance. 

\begin{remark}
The simulation reports exact floating-point values. In real deployments, deposits and rewards may be rounded to multiples of pence. This can be done by using integer constraints, rounding up to remain conservative, or rounding to the nearest value to stay close to the optimal solution.\\
\end{remark}

\begin{remark}
This simulation does not include the delay between announcing a reward and observing changes in user behaviour. In practice, users need time to notice new rewards and adjust their actions. The model also does not include the effect of higher deposits on the retail price of paper cup coffee, which may reduce demand for takeaway cups or increase the use of personal cups. In addition, we did not model how recycling habits may form over time and how stable habits could allow the deposit to decrease gradually. These behavioural delays, demand responses, and learning processes can be incorporated in future work using system dynamics and behavioural analysis tools. \\
\end{remark}

\section{Conclusion and Future Work}
\label{sec:con}

In this paper, we have improved a recent {DDRS} framework addressing the three gaps in digitalised recycling systems: bin congestion, decentralised reward allocation, and dynamic deposit tuning. For this purpose, we have proposed a stochastic load-balancing scheme based on Poisson signalling, an extended AIMD algorithm for splitting a fixed deposit across heterogeneous layers, and a PI-based controller that adjusts the deposit to meet a desired recycling rate.\\

Extensive simulations of the single and of the overall control problems we have addressed show that the decentralised assignment keeps bin levels balanced, the {AIMD} method converges to optimal rewards without revealing private behaviour functions, and the feedback loop reliably steers the recycling rate toward its target under realistic constraints. These results demonstrate that decentralised optimisation and feedback control can make {DDRS} architectures scalable, adaptive, and operationally effective.\\

While preliminary results are extremely positive, improving the modelling accuracy of the DDRS is our next objective. In particular, we are interested in integrating behavioural delays, more sophisticated demand response schemes, in addition to validation tests in real deployments, starting from a controlled environment such as a university campus.\\

\section*{Declaration of Interest}
The authors declare no competing financial interests or personal relationships that could have influenced the work reported in this paper.

\section*{Funding}
This work was supported by the UKRI underwrite of the European Union's Horizon-Europe Marie Sk{\l}odowska Curie Actions (MSCA), under grant agreement number 101073508. 

\section*{Data Availability Statement}
Data will be made available on request.

\bibliographystyle{elsarticle-num} 
\bibliography{zRef}

\end{document}